\documentclass[twocolumn,
showpacs,preprintnumbers,amsmath,amssymb]{revtex4}

\usepackage{graphicx}
\usepackage{amssymb,amsmath,amsthm}
\newtheorem{Thm}{Theorem}
\newtheorem{Cor}{Corollary}
\newtheorem{Lem}[Thm]{Lemma}
\newtheorem{Prop}[Thm]{Proposition}
\theoremstyle{definition}

\newcommand{\bra}[1]{{\left\langle #1 \right|}}
\newcommand{\ket}[1]{{\left| #1 \right\rangle}}

\newcommand{\M}{\mbox{$\mathbb M$}}
\newcommand{\T}{\mbox{$\mathrm{tr}$}}
\newcommand{\cH}{\mbox{$\mathcal{H}$}}
\newcommand{\C}{\mbox{$\mathbb{C}$}}
\newcommand{\E}{\mbox{$\mathcal{E}$}}

\begin{document}

\title{Zero-error classical capacity of qubit channels cannot be superactivated}

\author{Jeonghoon Park}
\affiliation{
 Department of Mathematics and Research Institute for Basic Sciences,
 Kyung Hee University, Seoul 130-701, Korea
}
\author{Soojoon Lee}\email{level@khu.ac.kr}
\affiliation{
 Department of Mathematics and Research Institute for Basic Sciences,
 Kyung Hee University, Seoul 130-701, Korea
}

\date{\today}

\begin{abstract}
It was shown [T.S.~Cubitt {\em et al.}, IEEE Trans. Inform. Theory {\bf 57}, 8114 (2011)] that
there exist quantum channels where a single use
cannot transmit classical information perfectly yet two uses can.
This phenomenon is called the superactivation of the zero-error classical capacity
which does not occur in classical channels.
In this paper, it is shown that qubit channels cannot generate the superactivation.
\end{abstract}

\maketitle

\section{Introduction}
Quantum information theory has provided us with many surprising and interesting results
that cannot be explained by classical information theory.
Among such results, there is a peculiar one, called {\em superactivation},
which can be described as the quantum phenomenon
to obtain a useful object with capacity
from several objects without capacity
by activating their hidden capabilities.

For example,
there are quantum states
which have nondistillable entanglement,
but from which distillable entanglement can be ingeniously extracted
by local operations and classical communication~\cite{SST},
there are two quantum channels
which have zero quantum capacity,
but whose joint quantum channel has nonzero quantum capacity~\cite{SY},
and there are also two quantum channels
which have no zero-error classical capacity,
but whose joint quantum channel has a positive zero-error classical capacity~\cite{CCH,Duan}
or a positive zero-error quantum capacity~\cite{CS}.
They are called the superactivation of bound entanglement,
the superactivation of quantum channel capacity,
the superactivation of the zero-error classical capacity of a quantum channel,
and the superactivation for quantum zero-error capacities, respectively.

Even though it has been known that
all bipartite entangled states including bound entangled states are useful
for quantum information processing~\cite{Masanes},
all bound entangled states do not seem to be superactivated.
Similarly, although it may be shown that
all quantum channels without a certain kind of capacity are useful in a sense,
this does not imply that the capacity can be superactivated.
Thus, the quantum effect called superactivation might be
such a rare phenomenon even in a quantum world that
it cannot be readily regarded as a quantum feature,
and hence
it could be important
to decide whether superactivation is feasible in a given situation.

We here take into account the zero-error classical capacity of a quantum channel,
which is the amount of classical information that can be perfectly transmitted
through the quantum channel.
In particular, a quantum channel $\E$ from system $A$ to system $B$
with a positive one-shot zero-error classical capacity
can be clearly expressed in a mathematical form as follows~\cite{CCH}:
\begin{equation}
\T[\E(\psi)^\dagger\E(\phi)] = 0
\label{eq:positive_capacity}
\end{equation}
for some pure states $\ket{\psi}$ and $\ket{\phi}$ in $\cH_A$.
Therefore, the superactivation of the one-shot zero-error classical capacity
can be mathematically described as follows:
The one-shot zero-error classical capacity of quantum channels $\E_1$, $\E_2, \ldots, \E_k$
can be {\em superactivated}
if and only if
\begin{equation}
\T[\E_j(\psi)^\dagger\E_j(\phi)] \neq 0,
\label{eq:zero_capacity}
\end{equation}
for all $1\le j \le k$ and all pure states $\ket{\psi}$ and $\ket{\phi}$ in $\cH_A$,
but there exist two pure states $\ket{\Psi}$ and $\ket{\Phi}$ in $\cH_A^{\otimes k}$ such that
\begin{equation}
\T[(\E_1\otimes \cdots \otimes \E_k)(\Psi)^\dagger(\E_1\otimes \cdots \otimes \E_k)(\Phi)] = 0.
\label{eq:positive_capacity_k}
\end{equation}
Here, Eq.~(\ref{eq:zero_capacity}) means that
no channel $\E_j$ has any one-shot zero-error classical capacity,
and Eq.~(\ref{eq:positive_capacity_k}) means that
the joint channel has a positive one-shot zero-error classical capacity.

We note that
there are quantum channels
whose one-shot zero-error classical capacity is superactivated~\cite{CCH},
but, for most quantum channels without the one-shot zero-error classical capacity,
their joint channels are unlikely to have a positive one-shot zero-error classical capacity.
On this account,
in order to figure out when or why the superactivation occurs,
it may be important
to learn what situation causes the superactivation, or
does not cause the superactivation.

In this paper, we present one situation
which cannot cause the superactivation of the one-shot zero-error classical capacity.
More precisely, we here show that
the one-shot zero-error classical capacity of
any finitely many qubit channels (with one qudit channel) cannot be superactivated,
which can be extended to the case of zero-error classical capacity for qubit channels.

\section{Choi-Jamio{\l}kowski Isomorphism and Superactivation}
In this section,
we introduce a necessary and sufficient condition for a positive one-shot zero-error classical capacity
based on the Choi-Jamio{\l}kowski isomorphism between matrices and linear operators~\cite{CJiso}.

We remark that
there is an isomorphism between $\mathcal{L}({M_n})$ and ${M_n}\otimes{M_n}$,
where ${M_n}$ is the space of $n \times n$ matrices and
$\mathcal{L}({M_n})$ is the space of all linear operators on $M_n$.
By the isomorphism, a quantum channel $\E$ corresponds to
$\sigma_{AA^{'}} = ({\mathcal{I}}_A \otimes \E_{A^{'}})(\omega_{AA^{'}})$,
where $\ket{\omega}_{AA^{'}}=\sum_{j}\ket{j}_A\ket{j}_{A'}$ and
$\omega_{AA^{'}}=\ket{\omega}_{AA^{'}}\bra{\omega}$.
This isomorphism is called the Choi-Jamio{\l}kowski (CJ) isomorphism,
and the matrix $\sigma_{AA^{'}}$ is called the CJ matrix of $\E$.

Let $\E$ be the channel,
${\E}^{*}$ be the dual map of $\E$
with respect to the Hilbert-Schmidt inner product
on the left-hand side of Eq.~(\ref{eq:positive_capacity}),
$\sigma$ be the CJ matrix of ${\E}^{*}\circ\E$, and
$S=\mathrm{supp}(\sigma)$ be the support of $\sigma$.
Then the following proposition can be obtained~\cite{CCH}.
\begin{Prop}\label{Prop:positive_S}
$\E$ has a positive one-shot zero-error classical capacity if and only if
there exist two pure states $\ket{\psi}$ and $\ket{\phi}$ in $\cH_A$
such that $\ket{\psi}\otimes\ket{\phi}\in S^\perp$.
\end{Prop}

By employing Proposition~\ref{Prop:positive_S},
the superactivation of the one-shot zero-error classical capacity
of two quantum channels $\E_1$ on system $A_1$ and $\E_2$ on system $A_2$
can be redescribed as follows:
For each $j$, let $\sigma_{j}$ be the CJ matrix of ${{\E}_{j}}^{*}\circ\E_{j}$
and $S_{j}^{A_{j}{A_{j}}'}=\mathrm{supp}(\sigma_j)$, and
let $S^{A_{1}A_{2}\textrm{-}{A_{1}}'{A_{2}}'}=
S_{1}^{A_{1}{A_{1}}'} \otimes S_{2}^{A_{2}{A_{2}}'}$.
Then the one-shot zero-error classical capacity
of two quantum channels $\E_1$ and $\E_2$ can be superactivated
if and only if,
for all pairs of pure states $\ket{\psi^j}$ and $\ket{\phi^j}$ in each $\cH_{A_j}$,
$\ket{\psi^j}\otimes\ket{\phi^j} \notin {S_{j}}^\perp$,
but there exist two pure states $\ket{\Psi}$ and $\ket{\Phi}\in \cH_{A_1}\otimes\cH_{A_2}$
such that $\ket{\Psi}\otimes\ket{\Phi}\in S^\perp$.

\section{Main Results}
In this section,
we show our main results
on the superactivation of the one-shot zero-error classical capacity.

\subsection{Main Lemma}
In this subsection,
we introduce the main lemma,
from which our main results can be straightforwardly derived.

Our main lemma is as follows:
\begin{Lem}\label{Lem:main}
For each $j$,
let $\E_j$ be a quantum channel on system $A_j$,
$\sigma_{j}$ be the CJ matrix of ${{\E}_{j}}^{*}\circ\E_{j}$, and
\begin{equation}
S_{j}^{A_{j}{A_{j}}'}=\mathrm{supp}(\sigma_j),
\label{eq:S_j_in_lemma}
\end{equation}
and let
\begin{equation}
S^{A_{1}A_{2}\textrm{-}{A_{1}}'{A_{2}}'}=
S_{1}^{A_{1}{A_{1}}'} \otimes S_{2}^{A_{2}{A_{2}}'}.
\label{eq:S_in_lemma}
\end{equation}
Assume that $\dim{S_{1}^{\perp}}\le 1$.
Then the one-shot zero-error classical capacity of the channels
cannot be superactivated, that is,
if 
 $\left(S_{j}^{A_{j}A_{j}}\right)^\perp$ does not contain
any product states with respect to partition ${A_{j}\textrm{-}A_{j}}$
for each $j=1, 2$,
then 
$\left(S^{A_{1}A_{2}\textrm{-}{A_{1}}'{A_{2}}'}\right)^{\perp}$
does not contain any product states
with respect to partition ${A_{1}A_{2}\textrm{-}{A_{1}}'{A_{2}}'}$,
either.
\end{Lem}

In order to prove Lemma~\ref{Lem:main},
we first consider the case that $S_{1}^{\perp}$ is one-dimensional.
Assume that $\dim{S_{1}^{\perp}}=1$.
Then we may let $\{\ket{\psi_1}\}$ be a basis for $S_{1}^{\perp}$,
$\{\ket{\psi_i}\}^{n^2}_{i=2}$ be a basis for $S_{1}$,
$\{\ket{\phi_i}\}^{k}_{i=1}$ be a basis for $S_{2}^{\perp}$, and
$\{\ket{\phi_i}\}^{m^2}_{i=k+1}$ be a basis for $S_{2}$,
where $n=\dim\cH_{A_1}$ and $m=\dim\cH_{A_2}$.
Thus, for any state $\ket{\Psi} \in (S_{1} \otimes S_{2})^{\perp}$,
it is clear that
\begin{eqnarray}
\ket{\Psi}
&=& \sum^{m^2}_{j=k+1} a_{j}\ket{\psi_1}\ket{\phi_j}
+ \sum^{n^2}_{i=1}\sum^{k}_{j=1} b_{ij}\ket{\psi_i}\ket{\phi_j}
\nonumber \\
&=& \ket{\psi_1} \left(\sum^{m^2}_{j=k+1} a_{j}\ket{\phi_j}\right)
+ \sum^{k}_{j=1} \left(\sum^{n^2}_{i=1} b_{ij}\ket{\psi_i}\right)\ket{\phi_j}
\nonumber \\
&=& \sum^{k+1}_{j=1} \ket{\tilde{\psi}_j}\ket{\tilde{\phi}_j},
\label{eq:dimS_1}
\end{eqnarray}
where $\ket{\tilde{\psi}_j}=\sum^{n^2}_{i=1} b_{ij}\ket{\psi_i}$
and $\ket{\tilde{\phi}_j}=\ket{\phi_j}$ for $1\le j\le k$,
$\ket{\tilde{\psi}_{k+1}}=\ket{\psi_1}$,
and $\ket{\tilde{\phi}_{k+1}}=\sum^{m^2}_{j=k+1} a_{j}\ket{\phi_j}$.

We now assume that $\dim{S_{1}^{\perp}}=0$.
Then we may let
$\{\ket{\psi_i}\}^{n^2}_{i=1}$ be a basis for $S_{1}=\cH_{A_1}\otimes\cH_{{A_1}'}$,
$\{\ket{\phi_i}\}^{k}_{i=1}$ be a basis for $S_{2}^{\perp}$, and
$\{\ket{\phi_i}\}^{m^2}_{i=k+1}$ be a basis for $S_{2}$.
Thus, for any state $\ket{\Psi} \in (S_{1} \otimes S_{2})^{\perp}$,
it is also obvious that
\begin{eqnarray}
\ket{\Psi}
&=&\sum^{n^2}_{i=1}\sum^{k}_{j=1} b_{ij}\ket{\psi_i}\ket{\phi_j}
\nonumber \\
&=&\sum^{k}_{j=1} \left(\sum^{n^2}_{i=1} b_{ij}\ket{\psi_i}\right)\ket{\phi_j}
\nonumber \\
&=&\sum^{k}_{j=1} \ket{\tilde{\psi}_j}\ket{\tilde{\phi}_j}.
\label{eq:dimS_0}
\end{eqnarray}
It follows from Eqs.~(\ref{eq:dimS_1}) and (\ref{eq:dimS_0}) that
the case of $\dim{S_{1}^{\perp}}=0$ is included in the case of $\dim{S_{1}^{\perp}}=1$.
Hence, it suffices to prove Lemma~\ref{Lem:main}
for the case that $S_{1}^{\perp}$ is one-dimensional.

For convenience of the proof of Lemma~\ref{Lem:main},
we rephrase the statement of Lemma~\ref{Lem:main} as its matrix-based version
by an isomorphism between states and matrices in the following proposition~\cite{CCH}.
\begin{Prop}\label{Prop:matrix_iso}
There exists an isomorphism $\M$
between (unnormalized) states in $\C^{d_A}\otimes\C^{d_B}$ and $d_A\times d_B$ matrices
defined as follows:
In the standard basis,
$\ket{\psi}_{AB} = \sum M_{ij}\ket{i}_A\ket{j}_B$ maps to $\M(\ket{\psi}_{AB})\equiv(M_{ij})$.
The isomorphism $\M$ has the following properties:
\begin{enumerate}
\item[(i)] Product states and entangled states correspond to
matrices of rank one and matrices of rank greater than one, respectively.
\item[(ii)] For any state $\ket{\Psi}_{11'\textrm{-}22'}=\sum^{k+1}_{i=1} \ket{\psi_i}\ket{\phi_i}$,
\begin{equation}
\M(\ket{\Psi}_{12\textrm{-}1'2'}) = \sum^{k+1}_{i=1} A^{i} \otimes B^{i},
\label{eq:M_Prop}
\end{equation}
where $A^{i}=\M(\ket{{\psi}_{i}}), B^{i}=\M(\ket{{\phi}_{i}})$.
\end{enumerate}
\end{Prop}
By applying Proposition~\ref{Prop:matrix_iso} to Eq.~(\ref{eq:dimS_1}),
Lemma~\ref{Lem:main} can be rewritten as the following lemma.
\begin{Lem}\label{Lem:main2}
Let $P$ and $Q$ be one-dimensional and $k$-dimensional
subspaces of $n \times n$ and $m \times m$ matrices
which have no matrices of rank one, respectively.
Let $\{A^{1}\}$ be a basis for $P$, $\{{B^{i}}\}^{k+1}_{i=2}$ be a basis for $Q$,
$B^1$ be an $m \times m$ matrix in $Q^{\perp}$, and $A^{j}$ be $n \times n$ matrices for $j\ge 2$.
Then $M={\sum}^{k+1}_{i=1} A^{i} \otimes B^{i}$ cannot be of rank one.
\end{Lem}
We now prove Lemma~\ref{Lem:main2}, which directly implies Lemma~\ref{Lem:main}.
\begin{proof}
Suppose that the matrix $M={\sum}^{k+1}_{i=1} A^{i} \otimes B^{i}$ is of rank one.
Then, by the singular value decomposition,
without loss of generality,
we may assume that
the matrix $A^{1}$ is a diagonal one with at least two positive diagonal entries,
since $A^1\in P$ is not of rank one.

We now consider an $m \times m$ submatrix $R_{st}$ of the matrix $M$
defined as
\begin{equation}
R_{st} \equiv \sum^{k+1}_{i=1}{(A^{i})_{st}}B^{i}.
\label{eq:R_st}
\end{equation}
Since the matrix $M$ is of rank one,
the submatrix $R_{st}$ must be the zero matrix or a rank-one matrix.
In particular, if $s \neq t$ then
the submatrix
\begin{equation}
R_{st} = \sum^{k+1}_{i=1}{(A^{i})_{st}}B^{i}
= \sum^{k+1}_{i=2}{(A^{i})_{st}}B^{i}
\label{eq:R_st_in_Q}
\end{equation}
is contained in $Q$,
which has no matrices of rank one,
since $A^1$ is diagonal and $\{{B^{i}}\}^{k+1}_{i=2}$ is a basis for $Q$.
Hence we obtain that the submatrix $R_{st}$ must be the zero matrix for $s\neq t$,
and $A^i$ is diagonal for each $2\le i\le k+1$.

We now take into account the submatrix $R_{st}$ for $s=t$, that is, $R_{ss}$.
We first assume that $B^1$ is the zero matrix.
Then, similar to Eq.~(\ref{eq:R_st_in_Q}), the submatrix
\begin{equation}
R_{ss}=\sum^{k+1}_{i=2}{(A^{i})_{ss}}B^{i}
\label{eq:R_ss_in_Q}
\end{equation}
is contained in $Q$,
and hence $R_{ss}$ cannot be of rank one.
Thus, for all $s$, $R_{ss}$ must be the zero matrix,
and $(A^i)_{ss}$ is zero for each $2\le i\le k+1$, that is,
$A^i$ is the zero matrix for each $2\le i\le k+1$.
Since $B^1$ is the zero matrix,
this implies that the matrix $M$ is the zero matrix,
which is a contradiction.

We now assume that $B^1$ is nonzero.
Then it follows that
when $(A^1)_{ss} \neq 0$,
the submatrix
\begin{equation}
R_{ss}=\sum_{i=1}^{k+1}(A^i)_{ss} B^i
\label{eq:R_ss}
\end{equation}
cannot be the zero matrix,
and thus it must be of rank one.
Since $A^{1}$ has at least two positive diagonal entries,
there exist two distinct submatrices $R_{\alpha\alpha}$ and $R_{\beta\beta}$ of rank one.
The matrix $M$ must have rank greater than one,
since $R_{\alpha\alpha}$ and $R_{\beta\beta}$ do not share any entries of $M$
and both $R_{\alpha\beta}$ and $R_{\beta\alpha}$ are the zero matrices.
This leads to a contradiction.

Therefore, the matrix $M$ cannot be a rank-one matrix.
\end{proof}

\subsection{Main Theorem and Main Corollaries}
In this subsection, we phrase our main theorem and corollaries.

From Lemma~\ref{Lem:main2} (or Lemma~\ref{Lem:main}),
we clearly obtain our main theorem.
\begin{Thm}\label{Thm:main}
For each $j$,
let $\E_j$ be a quantum channel on system $A_j$,
$\sigma_{j}$ be the CJ matrix of ${{\E}_{j}}^{*}\circ\E_{j}$,
\begin{equation}
S_{j}^{A_{j}{A_{j}}'}=\mathrm{supp}(\sigma_j),
\label{eq:S_j_in_theorem}
\end{equation}
and
\begin{equation}
S^{A_{1}A_{2}\textrm{-}{A_{1}}'{A_{2}}'}=
S_{1}^{A_{1}{A_{1}}'} \otimes S_{2}^{A_{2}{A_{2}}'}.
\label{eq:S_in_theorem}
\end{equation}
Assume that $\dim{S_{1}^{\perp}}\le 1$.
Then the one-shot zero-error classical capacity of the quantum channels $\E_j$
cannot be superactivated.
\end{Thm}

In order to consider the quantum channels on the two-dimensional quantum system,
that is, the qubit channels,
we use the following lemma~\cite{CMW}.
\begin{Lem}\label{Lem:dim}
For any subspace $S$ of $\C^{d_A}\otimes\C^{d_B}$
whose states all have at least a Schmidt number of $r$,
the maximum dimension of $S$ is $({d_A}-r+1)({d_B}-r+1)$.
\end{Lem}
It follows from Lemma~\ref{Lem:dim} that
if a quantum channel $\E_j$ on system $A_j$ is a qubit channel
and $S_j$ is the subspace induced by channel $\E_j$
then  $\dim{S_{j}^{\perp}}$ is 0 or 1,
since $\dim{\cH_{A_j}}=2$.
Hence, we can readily obtain the following main corollaries.
\begin{Cor}\label{Cor:main1}
The one-shot zero-error classical capacity of two quantum channels
including at least one qubit channels
cannot be superactivated.
\end{Cor}

By the induction on the number of quantum channels,
we clearly have the following corollary.
\begin{Cor}\label{Cor:main2}
No finitely many qubit channels (with one qudit channel)
can cause the superactivation of the one-shot zero-error classical capacity.
\end{Cor}

We remark that
nonsuperactivation of the one-shot zero-error classical capacity for qubit channels
implies non-superactivation of zero-error classical capacity
by their definitions.
As a consequence,
our result for the one-shot zero-error classical capacity of qubit channels
can be extended to the case of zero-error classical capacity.

\section{Conclusions}
We have investigated whether the superactivation of the zero-error classical capacity
arises in quantum channels on low-dimensional quantum systems,
and have presented a necessary condition for the superactivation of
the one-shot zero-error classical capacity,
which can be reduced to the application of the qubit channels.
It has been shown that
the one-shot zero-error classical capacity of two quantum channels
including at least one qubit channels
cannot be superactivated,
and no finitely many qubit channels (with one qudit channel)
can cause the superactivation of the one-shot zero-error classical capacity.

We note that
there have never been
any examples of the superactivation for channel capacities in the literature,
when the underlying space is a two-dimensional one.
Therefore, our results could be applied to the superactivation,
and could be generalized to the conclusion that
the qubit systems cannot cause the superactivation for any channel capacities.

\acknowledgments{
This work was supported
by the IT R\&D program of the Ministry of Knowledge Economy
[Development of Privacy Enhancing Cryptography on Ubiquitous Computing Environment]
and Basic Science Research Program
through the National Research Foundation of Korea (NRF)
funded by the Ministry of Education, Science and Technology (Grant No.~2009-0076578).
}

\end{document}